\documentclass{article}
\usepackage[affil-it]{authblk}
\usepackage[english]{babel}

\usepackage[letterpaper,top=2cm,bottom=2cm,left=3cm,right=3cm,marginparwidth=1.75cm]{geometry}

\usepackage{amsmath}
\usepackage{graphicx}
\usepackage{color,xcolor}

\usepackage[colorlinks=true, allcolors=blue]{hyperref}
\usepackage{float}
\usepackage{xurl}

\title{A hypothesis-free bridging of disease dynamics and non-pharmaceutical policies
}
\author[1,2]{Xiunan Wang}
\author[1]{Hao Wang\footnote{The corresponding author. Email: hao8@ualberta.ca}}
\author[3]{Pouria Ramazi} 
\author[1,4]{Kyeongah Nah} 
\author[1,5]{Mark Lewis}

\affil[1]{Department of Mathematical and Statistical Sciences, University of Alberta, Edmonton, AB T6G 2G1, Canada}
\affil[2]{Department of Mathematics, University of Tennessee at Chattanooga, Chattanooga, TN 37403, USA
}
\affil[3]{Department of Mathematics and Statistics, Brock University, St. Catharines, ON L2S 3A1, Canada}
\affil[4]{National Institute for Mathematical Sciences, Daejeon 34047, Korea}
\affil[5]{Department of Biological Sciences, University of Alberta, Edmonton, AB T6G 2G1, Canada}

\begin{document}
\maketitle

\begin{abstract}
Accurate prediction of the number of daily or weekly confirmed cases of COVID-19 is critical to the control of the pandemic. 
Existing mechanistic models nicely capture the disease dynamics. 
However, to forecast the future, they require the transmission rate to be known, limiting their prediction power. 
Typically, a hypothesis is made on the form of the transmission rate with respect to time. 
Yet the real form is too complex to be mechanistically modeled due to the unknown dynamics of many influential factors. 
We tackle this problem by using a hypothesis-free machine-learning algorithm to estimate the transmission rate from data on non-pharmaceutical policies, and in turn forecast the confirmed cases using a mechanistic disease model. 
More specifically, we build a hybrid model consisting of a mechanistic ordinary differential equation (ODE) model and a gradient boosting model (GBM).
To calibrate the parameters, we develop an ``inverse method'' that obtains the transmission rate inversely from the other variables in the ODE model and then feed it into the GBM to connect with the policy data. 
The resulting model forecasted the number of daily confirmed cases up to 35 days in the future in the United States with an averaged mean absolute percentage error of 27\%. 
It can identify the most informative predictive variables, which can be helpful in designing improved forecasters as well as informing policymakers. 
\end{abstract}

%\section*{Title candidates:}

%1) A novel two-step method for connecting a dynamical system for Covid-19 with machine learning\\

%\noindent 2) Accurate prediction of COVID-19 dynamics by linking epidemiological dynamics to machine learning\\

%\noindent 3) A hypothesis-free approach for connecting disease dynamics to covariates:  inverse method plus machine learning\\

%\noindent 4) A hypothesis-free approach for connecting disease dynamics to policies:  inverse method plus machine learning\\

%\noindent 5) Predicting infectious diseases by coupling disease dynamics to covariates 1: pharmaceutical policies\\

%\noindent 6) A hypothesis-free approach for connecting disease dynamics to non-pharmacological policies:  inverse method plus machine learning \\

%\noindent 7) From policy to prediction: a hypothesis-free method for determining disease dynamics\\

%\noindent 8) From policy to prediction: a hypothesis-free method to forecast disease dynamics

\section{Introduction}\label{Introduction}

The world has experienced a devastating pandemic of COVID-19, a novel coronavirus disease caused by SARS-CoV-2. As of November 16, 2021, the COVID-19 pandemic is still affecting $224$ countries and territories, causing about $254,901,115$ cases and $5,127,051$ deaths worldwide \cite{worldometer}. The first case in the US was reported on January 23, 2020 \cite{Wikipedia} and the first death in the US was reported on February 29, 2020 \cite{worldometer}. The confirmed cases and deaths kept increasing in the US in 2020, making it the epicenter. 
In the beginning several months of the pandemic, pharmaceutical interventions such as vaccination and drugs are not available, and containing the spread of SARS-CoV-2 largely depends on government policies including school closing, workplace closing, cancellation of public events, restrictions on gatherings, public transport closing, stay at home requirements, international travel controls, public information campaigns, testing, contact tracing, facial coverings, protection of elderly people, etc. \cite{owidcoronavirus}. 
Most of these policies directly affect human mobility which further influence the transmission of the virus. 
Revealing the quantitative relationship between the transmission rate and policies and human mobilities is critical in forecasting the pandemic.
%, and the methods can be modified to predict the COVID-19 dynamics when vaccination or antiviral drugs are available.

There has been an overwhelming number of research papers about the transmission dynamics of COVID-19 (e.g., \cite{coletti2021data,mukandavire2020quantifying,sun2020forecasting,liu2020covid,ihme2020modeling,chang2021mobility,calvetti2020metapopulation, ramazi2021accurate}). 
Nonetheless, intuitive modeling and accurate forecasting of the spread of COVID-19 remain a challenge. 
On one hand, the traditional epidemiological models are fully mechanistic and intuitive.
They nicely capture the disease spread yet heavily rely on the transmission rate parameter which in turn depends on variables such as preventive policies and human mobility, whose relation to the disease dynamics is too complex to be accurately modelled mechanistically. 
Therefore, to forecast the future, the transmission rate is considered constant or piecewise linear, or some restrictive hypothesis is made about its future values.
The mechanistic models are, thus, often not competent enough in prediction. 
On the other hand, the data-based machine learning models are powerful in prediction but typically non-intuitive, and perhaps less reliable, especially if trained with few data instances. 
We bridge the gap by developing a hybrid model combining a compartmental epidemiological model that captures the disease spread with a time-varying transmission rate and a machine-learning model that links the transmission rate to data on preventive policies whose future values are known \emph{a priori}. 
The epidemiological model consists of an ordinary differential equations (ODE) and a machine-learning algorithm - a gradient boosting model (GBM). 
We use part of the available data to \emph{train} the GBM, by first, developing an ``inverse method'' that estimates the values of the transmission rate from the other variables of the ODE, and next, fitting the estimated transmission rate values to the policy data using the GBM. 
The trained hybrid model can then be used to generate predictions of the number of daily confirmed cases by using the future values of the preventive policies to estimate the transmission rate by the GBM and in turn, the daily cases using the ODE. 
To examine the role of human mobility on the disease spread, we run a separate series of simulations where in addition to the preventive policies, human mobility data is used to estimate the transmission rate. 
We apply the model to the case study of COVID-19 in the United States, and then find those variables whose inclusion improved the model performance most. 

The rest of the paper is organized as follows. 
Section \ref{DataSection} explains the data used in this study. In Section \ref{ODEsection}, we develop the compartmental epidemiological model for COVID-19 and introduce the inverse method to estimate the transmission rate. In Section \ref{MachineLearningPolicySection}, we formulate the generalized boosting model, show the training and testing results and make predictions of daily confirmed cases using the ordinary differential equation (ODE) model. We also explore the relative importance of each variable in training the model. In Section \ref{MachineLearningMobilitySection}, we investigate the prediction performance when mobility and part of the policies are additionally included as the predictor variables.
Section \ref{Discussion} provides a brief summary of the method and results as well as suggestions for future work.

\section{Data Availability}\label{DataSection}
The data used in this study includes the total number of daily confirmed cases of COVID-19 in the US and policy indices in each state collected from the official website of the flagship project {\it Our World in Data} of Global Change Data Lab \cite{owidcoronavirus} (\url{https://ourworldindata.org/coronavirus}), the six categories of human mobility data in the US from the official website of Google \cite{Google} (\url{https://www.google.com/covid19/mobility/}), and deaths, recovered and active cases in the US from the worldometer website \cite{worldometer} (\url{https://www.worldometers.info/coronavirus/country/us/}), on a daily basis from April 4, 2020 to December 19, 2020.

\begin{figure}[hp]
\centering
\includegraphics[width=0.49\textwidth]{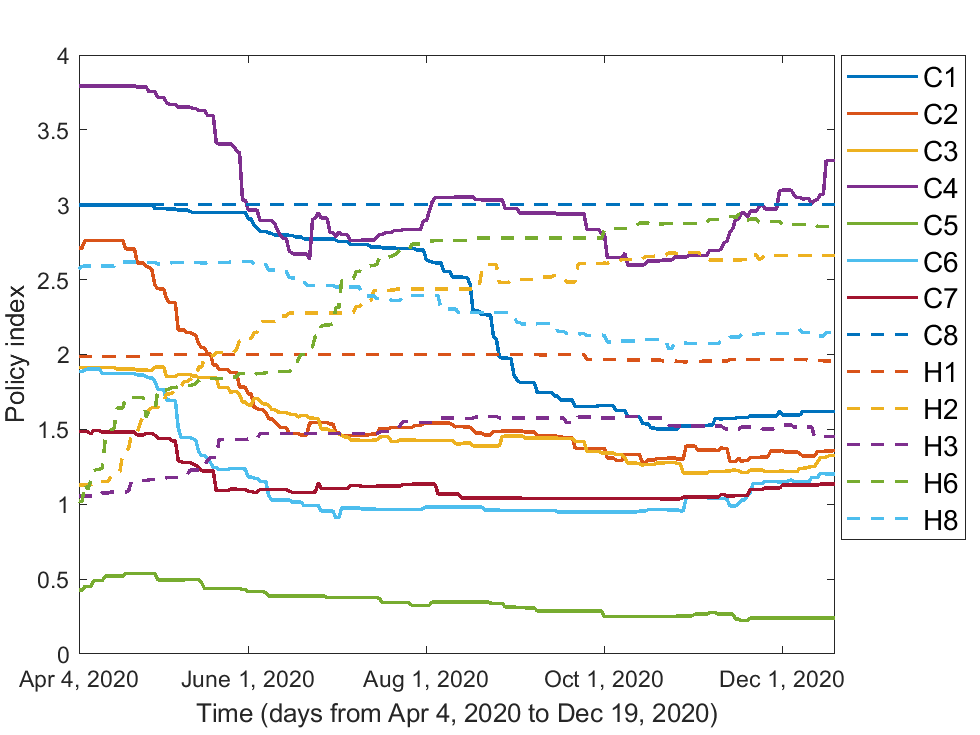}
\includegraphics[width=0.49\textwidth]{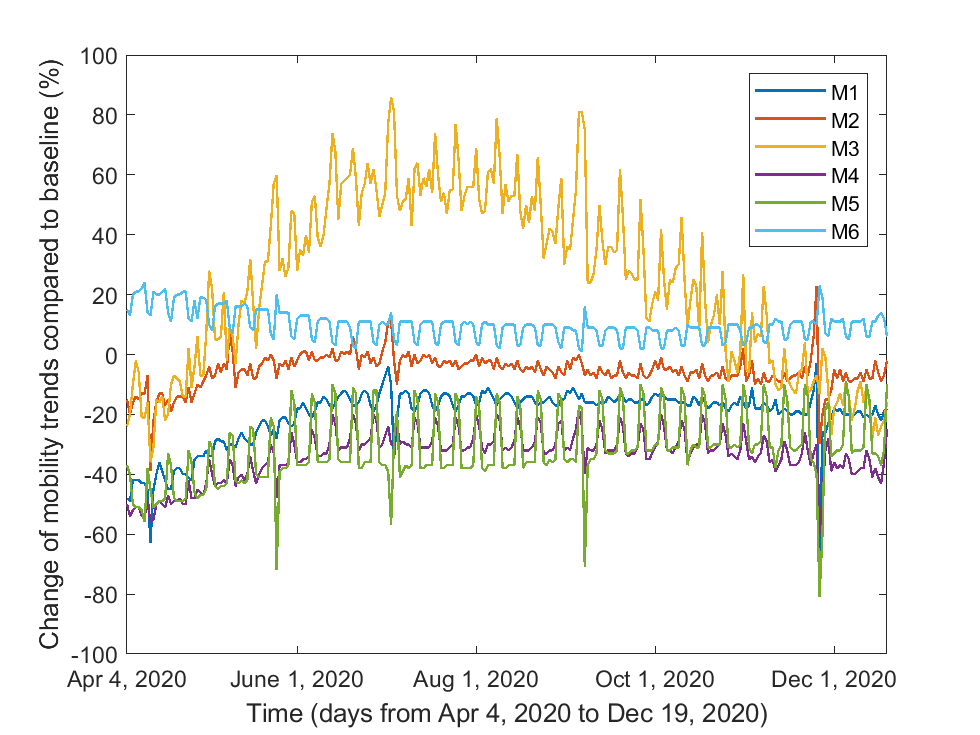}
\caption{\label{PolicyFig} Policy and mobility data in the US from Apr 4, 2020 to Dec 19, 2020.}
\end{figure}

We obtain the time-series indices for school closing (denoted by C1), workplace closing (C2), cancel public events (C3), restrictions on gatherings (C4), close public transport (C5), stay at home requirements (C6), restrictions on internal movement (C7), international travel controls (C8), public information campaigns (H1), testing policies (H2), contact tracing (H3), facial coverings (H6), and protection of elderly people (H8) in the US by taking an average of the corresponding policy indices over all the 50 US states as well as Washington D.C., weighted by their populations. Here the policies beginning with ``C" represent containment policies whereas those beginning with ``H" represent health policies. The emergency investment in healthcare (H4) and investment in vaccines (H5) are not available. Since we focus on the pre-vaccination case in this paper, we do not take into account the vaccination delivery policy (H7) either.
Human mobility data include changes of mobility trends ($\%$) in retail and recreation (M1), grocery and pharmacy (M2), parks (M3), transit stations (M4), workplaces (M5), and residential (M6), compared to the baseline level (0). These policy and mobility data are shown in Figure \ref{PolicyFig}.

\section{Mechanistic Model and Inverse Method
}\label{ODEsection}
We use a susceptible-exposed-infectious-recovered framework to model the transmission dynamics of COVID-19. The model divides the human population into five compartments: the susceptible (denoted by $S$), the exposed  ($E$), the symptomatic infected ($I$), the asymptomatic infected ($A$), and the recovered individuals ($R$). Our SEIAR model is described by the following system of differential equations:

\begin{equation}
		\label{pre_model}
		\aligned
		\frac{\mathrm{d}S(t)}{\mathrm{d}t} =& 
		-\frac{\beta(t)S(t)(I(t)+\theta_EE(t)+\theta_AA(t))}{N}, \\
		\frac{\mathrm{d}E(t)}{\mathrm{d}t} =& 
		\frac{\beta(t)S(t)(I(t)+\theta_EE(t)+\theta_AA(t))}{N}-\delta E(t), \\
		\frac{\mathrm{d}I(t)}{\mathrm{d}t} =& 
		(1-p)\delta E(t)-(\mu(t)+r_I)I(t), \\
		\frac{\mathrm{d}A(t)}{\mathrm{d}t} =& 
		p\delta E(t)-r_AA(t), \\
		\frac{\mathrm{d}R(t)}{\mathrm{d}t} =& 
		r_II(t)+r_AA(t). \\
		\endaligned
\end{equation}
The susceptible individuals enter the incubation period if they are infected with SARS-CoV-2. The incubation period has an average duration of $1/\delta$ days. Upon the incubation period ends, the exposed individuals enter either the symptomatic infected compartment (I) or the asymptomatic infected compartment (A), depending on whether symptoms occur or not. We assume that a proportion $p$ of all the infectives are asymptomatic and hence the symptomatic infections account for a proportion of $1-p$. The transmission rate is $\beta(t)$. As exposed individuals and asymptomatic infected individuals can also spread the virus at reduced probabilities compared to symptomatic infected individuals \cite{zhang2020estimated}, we assume that the relative transmissibility of exposed and asymptomatic infected individuals are $\theta_E$ and $\theta_A$, respectively ($0\le\theta_E\le1$, $0\le\theta_A\le1$). The disease induced death rate is $\mu(t)$. It takes an average of $1/r_I$ days and $1/r_A$ days for symptomatic and asymptomatic infected individuals to recover, respectively.

%Step 1: Estimate parameter values (obtained from references, estimate time-varying death rate from death data, use inverse method to estimate time-varying transmission rate).\\

We obtain the values of the constant parameters from the literature. The total US population $N$ is taken as 331,449,281 which is estimated on April 1, 2020 by US Census Bureau \cite{USCensusBureau}. The incubation period could vary greatly among patients. The current official estimated range for the incubation period is $2$ to $14$ days. However, more recent reports show that the incubation period can extend beyond $14$ days (\url{https://www.news-medical.net/news/20201025/COVID-19-incubation-period-potentially-much-longer-than-previously-thought.aspx}
). We take $\delta=1/14$ per day. The time to recover from COVID-19 may vary from $1.5$ to $30$ days among different patients \cite{kumar2021time}, depending on their infection severity, overall health and age. We assume the average recovery period for both symptomatic and asymptomatic infected individuals is $14$ days, which leads to $r_I=r_A=1/14$ per day. Asymptomatic infections contribute substantially to community transmission together with presymptomatic ones. Even if asymptomatic infections transmit poorly, presymptomatic and asymptomatic cases together comprise at least $50\%$ of the force of infection \cite{Rahul2021}. We set $p=0.7$ to represent that approximately $70\%$ of the infections are asymptomatic in our model. We estimate the relative transmissibilities of exposed and asymptomatic infected individuals as
$\theta_E=0.1$ and $\theta_A=0.5$, respectively. 
The values and interpretations of all constant parameters are given in Table 1.\\

The time-varying death rate is estimated using the following formula where $\mu[k]$ represents the disease induced death rate of symptomatic infected individuals on day $k$:
$$\mu[k]=\frac{\text{\# new deaths on day}\ k}{\# \text{currently infected individuals on day}\ k}.$$

Motivated by \cite{Kong2015,Pollicott2012}, we create an inverse method to estimate the time-varying transmission rate. The starting point is to derive the time series $E(t)$ by utilizing the notification data. The real incidence data will be between $\delta E(t)$ and $(1-p)\delta E(t)$, but most asymptomatic individuals are not tested due to unawareness of their infections. Although some special individuals such as sports players or frontline health workers may be forced to be tested, this accounts for a tiny portion of the total population. Some regions in China (e.g., Wuhan, Shenyang, Guangzhou) tested everyone once several new cases were reported locally. However, this never happened in the US. Hence we use the values of $(1-p)\delta E(t)$ as an approximation of the notification data. 

We use $S[k]$, $E[k]$, $I[k]$, $A[k]$ and $R[k]$ to represent the values of variables in model \eqref{pre_model} and $y[k]$ to be the notification data on the $k$-th day of study. In addition, we use $D[k]$ to represent the cumulative death number on the $k$-th day. Then we have
$$E[k]=\frac{y[k]}{(1-p)\delta},\quad k=1,2,...,K,$$
where $K$ is the length of the vector of the notification data.
We estimate the initial values $I[1]$, $R[1]$ and $D[1]$ from reporting data \cite{owidcoronavirus,worldometer}: $I[1]=21637$, $R[1]=14813$, $D[1]=10595$. Moreover, we assume that $A[1]=2I[1]$ considering that most infected people are asymptomatic \cite{Rahul2021}. Then $S[1]=N-E[1]-I[1]-A[1]-R[1]-D[1]$.
It follows that
\begin{align*}
    I[k]&=I[k-1]+(1-p)\delta E[k-1]-(\mu[k-1]+r_I)I[k-1],\\
    A[k]&=A[k-1]+p\delta E[k-1]-r_A A[k-1],\\
    R[k]&=R[k-1]+r_I I[k-1]+r_A A[k-1],\\
    D[k]&=D[k-1]+\mu[k-1]I[k-1],\\
    S[k]&=N-E[k]-I[k]-A[k]-R[k]-D[k],\\
    \beta[k-1]&=-\frac{N(S[k]-S[k-1])}{(S[k-1](\theta_E E[k-1]+\theta_A A[k-1]+I[k-1]))},
\end{align*}
for $k=2,3,...K$.  Approximately, we have $\beta[K]\approx\beta[K-1]$.
The idea is that once we get the time series values of $E(t)$, we are able to obtain the time series values of $I(t)$, $A(t)$, $R(t)$, and hence, $S(t)$. Then from the first equation of system \eqref{pre_model}, we can solve for $\beta(t)$.
Note that the inverse method used in this study is different from that in \cite{Kong2015,Pollicott2012}, although the essential idea is similar, that is, to solve for the transmission rate inversely. 

%\noindent Step 2: Train the GBM model for different lengths of days (from 105 days to 231 days by 7 days, starting from April 4, 2020) by fitting log(beta) to mobility data and policy data.\\

%\noindent Step 3: Test the GBM model for 35 days right after each train duration by using the predict function in R.\\

%\noindent Step 4: Plot simulated curve of $(1-p)\delta E(t)$ of the SEIAR model by using the trained and tested values of transmission rate and compare with notification data. Compute the MAE and MAPE corresponding to different lengths of training durations, and then obtain the averaged MAE and MAPE for the test results of each of the above GBM model. \\

\section{Machine Learning and Prediction}\label{MachineLearningPolicySection}
Human mobility can affect the transmission rate, and policies from the government may affect human mobility. Therefore, the transmission rate can be indirectly affected by the policies. Indeed, some policies such as facial coverings may even directly affect the transmission rate.  We use a GBM to estimate the transmission rate from the policy predictor variables: C1$\sim$C8, H1, H2, H3, H6, H8.

Having estimated the transmission rate in Section \ref{ODEsection}, we can fit $\log(\beta(t))$ with mobility and policy data using the GBM.
We partition the data into a \emph{training dataset}, used to calibrate the parameters, and a \emph{testing dataset}, used to test the model performance in making predictions.
The partitioning should be temporal: 
Since the model is supposed to make predictions in the future, it should be tested on a dataset that is ``in the future'' compared to the dataset used for estimating the model parameters, where the values of the number of confirmed cases are unavailable.
More specifically, the data instances from time $0$ to $T$ are used for training and from time $T+1$ to $T+T'$, for some $T,T'>0$, are used for testing the model. 
We may, otherwise, obtain misleadingly high performance results \cite{ramazi2021accurate,ramazi2021Predicting}.

We fix the start date of the training at April 4, 2020 and let the training duration increase from 105 days to 224 days by a step size of 7 days (see Table \ref{preVac}). The training dataset consists of the transmission rate on each day obtained by the inverse method as the response variable and all the 13 types of policy data C1$\sim$C8, H1, H2, H3, H6 and H8 on each day as predictor variables for the GBM. 
We fix the test duration at 35 days right after each training duration (see Table \ref{preVac}).
The trained GBMs will predict the transmission rate based on the policy data provided during the test duration.
The \texttt{gbm} package and the \texttt{predict} function in \texttt{R} are used.

Then we can plot the curve of $(1-p)\delta E(t)$ of the SEIAR model \eqref{pre_model} by using the time series of trained and tested daily transmission rates to compare with notification data of COVID-19 confirmed cases. 
To evaluate the fitting results, we use the mean absolute error (MAE) and the mean absolute percentage error (MAPE) to compute the differences between the transmission rates predicted by the GBM and those obtained by the inverse method as well as the differences between the predicted and actual numbers of daily COVID-19 confirmed cases.
The formulas of MAE and MAPE are given by
$$
\text{MAE}=\frac{1}{n}\sum_{i=1}^n|y_i-x_i|,\quad
\text{MAPE}=\frac{1}{n}\sum_{i=1}^n\left|\frac{y_i-x_i}{x_i}\right|,
$$
where $x_i$ is the $i$-th component of the vector of actual values, $y_i$ is the $i$-th component of the vector of prediction values, and
$n$ is the total number of data instances.

Gradient boosting trains models gradually, additively, and sequentially by minimizing the loss function via the number of trees. Alongside with the number of trees, the other parameters including the distribution of response variable, the stochastic gradient descent, the learning rate, the depth of interaction, and the minimum number of observations allowed in the trees’ terminal nodes can all directly affect the performance of the model \cite{Mayr2015,ZhangNPL2019}. We select the parameter values for GBM based on the averaged MAE and MAPE over the different training durations in Table \ref{preVac}. 
We apply the \texttt{summary} function with the default method of relative influence in R to investigate the variable importance in training the model. 

After trying different combinations of the GBM parameters, we decide to employ 1000 trees with a Gaussian distribution of the response variable, 0.9 stochastic gradient descent, 0.01 learning rate, 30 depth of interaction and a minimum number of 10 observations allowed in the trees' terminal nodes, which results in a smaller averaged MAE and MAPE for predictions based on the various training durations in Table \ref{preVac}.

The prediction performance of the
GBM for the daily confirmed cases of
COVID-19 is summarized in Tables \ref{allMAEMAPE} and \ref{averMAEMAPE}. 
We can see that small MAE and MAPE are obtained when the GBM
is trained for 126 days, 147 days, 154 days and 175 days. The corresponding training and testing (prediction) results of the transmission rate together with those of confirmed cases are presented in Figure \ref{fit01}, supplementary Figures \ref{fit02}, \ref{fit03} and \ref{fit04} (see Appendix). 
The trained transmission rates (i.e., the orange curves in the left panels of these figures) generally fit well with the ones obtained from the inverse method (i.e., the blue curves in the left panels of these figures). However, the tested transmission rate (i.e., the yellow curves in the left panels of these figures) do not fit well with the peaks or troughs of the blue curves of the transmission rate. In the right panels of Figure \ref{fit01} and supplementary Figures \ref{fit02}, \ref{fit03}, \ref{fit04}, the orange curves (i.e., trained part) fit almost perfectly with the real notification data of confirmed cases. The yellow curve of prediction in the right panel of Figure \ref{fit01} also fits well with the blue circles of real data, with the MAPE equal to $4.92\%$. In the right panel of supplementary Figures \ref{fit02} and \ref{fit03}, the yellow prediction curve does not show a good fitting with the local minimum point around September 11 although the MAPE is as small as $6.47\%$ and $6.45\%$, respectively. 

\begin{figure}[ht]
\centering
\includegraphics[width=0.6\textwidth]{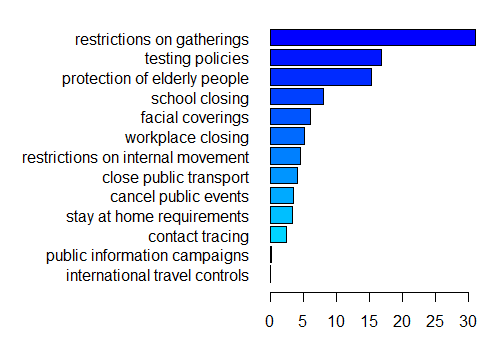}
\caption{\label{figPre126P} Relative influence of policy variables when trained for 126 days from Apr 4, 2020 to Aug 7, 2020.}
\end{figure}

The relative influence of a variable in a single tree is the sum of the empirical improvement by splitting on the variable at those points. Friedman extended it to boosting models by averaging the relative influence of each variable across all the trees generated by the boosting algorithm \cite{Friedman2001}.
The relative influence of mobility and policy variables for the GBM based on the different training durations of $126$ days, $147$ days, $154$ days and $175$ days are shown in Table, \ref{tablePre126P},
supplementary Tables \ref{tablePre147P},
\ref{tablePre154P} and \ref{tablePre175P}, respectively. 
Among these policies, restrictions on gatherings always have the highest weight of relative influence which is as large as $42.46\%$ when trained for $154$ days. Other important predictors are testing policies, facial coverings, school closing, protection of elderly people and workplace closing. Public information campaigns and international travel controls are the least important policies with a weight of at most $0.43\%$ for public information campaigns when trained for 147 days and zero influence from international travel controls (see Table \ref{tablePre126P}, supplementary Tables \ref{tablePre147P}, \ref{tablePre154P} and \ref{tablePre175P}).
As can be seen from Figure \ref{figPre126P} and supplementary Figures \ref{figPre147P}, \ref{figPre154P} and \ref{figPre175P}, the rankings of the relative influence of some policy variables have changed when trained for different lengths of days. 

\section{Machine Learning with Policy and Mobility Data}\label{MachineLearningMobilitySection}
While fitting the transmission rate with policy data is helpful for prediction, it would be interesting to see how the transmission rate can be affected by mobility as well since human mobility is considered to have direct impact on the transmission rate. Among all the policies that we have investigated in Section \ref{MachineLearningPolicySection}, testing policies (H2), contact tracing (H3) and facial coverings (H6) normally do not affect human mobility. Thus, it is reasonable to set H2, H3, H6 and mobility variables M1$\sim$M6 as the predictor variables and to keep all the mobility variables unchanged while changing some of these policies when we explore the effects of these three policies on the transmission rate.
In this section, We use GBM to connect the transmission rate with mobility data in the presence or absence of policy data. We perform two GBMs with different predictor variables: one involves the mobility variables M1$\sim$M6 only; the other consists of both the mobility variables M1$\sim$M6 and the policy variables H2, H3, H6.
	
We use the same values of parameters as those in Section \ref{MachineLearningPolicySection}, train the two GBMs for different training durations increasing from 105 days to 224 days by 7 days, and test the models for 35 days following each training duration (see Table \ref{preVac}). The training dataset consists of the transmission rate on each day obtained by the inverse method as the response variable and all the six types of mobility data M1$\sim$M6 on each day as predictor variables for both GBMs. Additionally, the training dataset of the GBM involving both mobility and poilcy variables includes the three types of policy data H2, H3 and H6 on each day as predictor variables as well. 
The trained GBMs will give a prediction for the transmission rate based on the mobility and/or policy data provided during the test duration.
Then we can plot the curve of $(1-p)\delta E(t)$ of the SEIAR model \eqref{pre_model} by using the time series of trained and tested daily transmission rates to compare with notification data of confirmed cases. 

\begin{figure}[ht]
\centering
\includegraphics[width=0.49\textwidth]{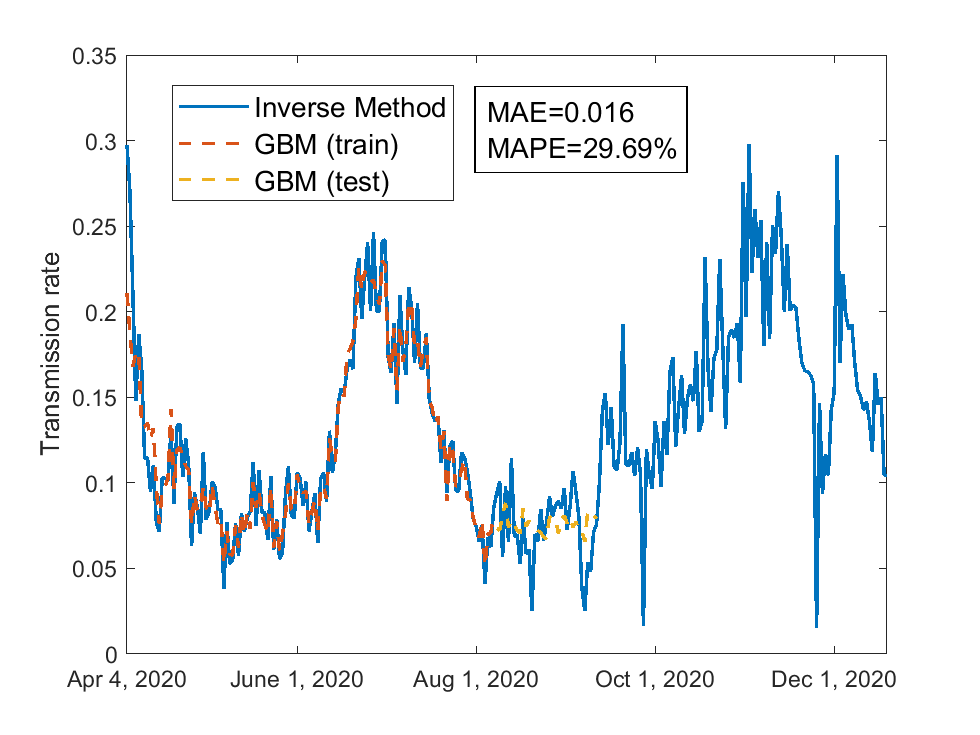}
\includegraphics[width=0.49\textwidth]{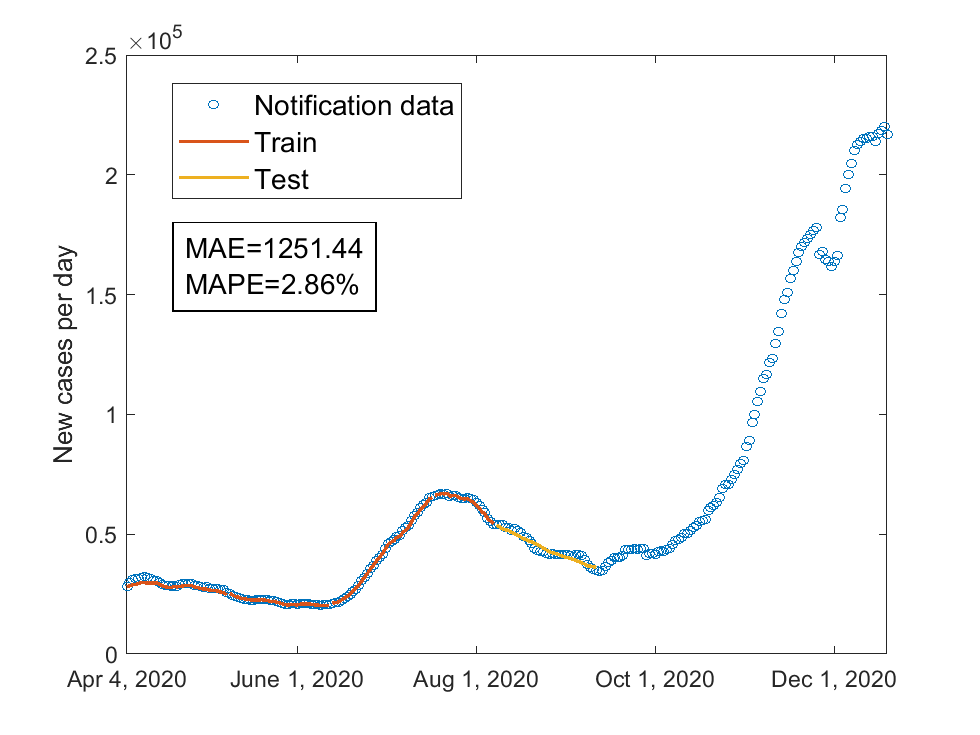}
\caption{\label{fit1} Using mobility data M1$\sim$M6 and policy data H2, H3, H6, train 126 days from Apr 4, 2020 to Aug 7, 2020; test 35 days from Aug 8, 2020 to Sept 11, 2020.}
\end{figure}

\begin{figure}[ht]
\centering
\includegraphics[width=0.49\textwidth]{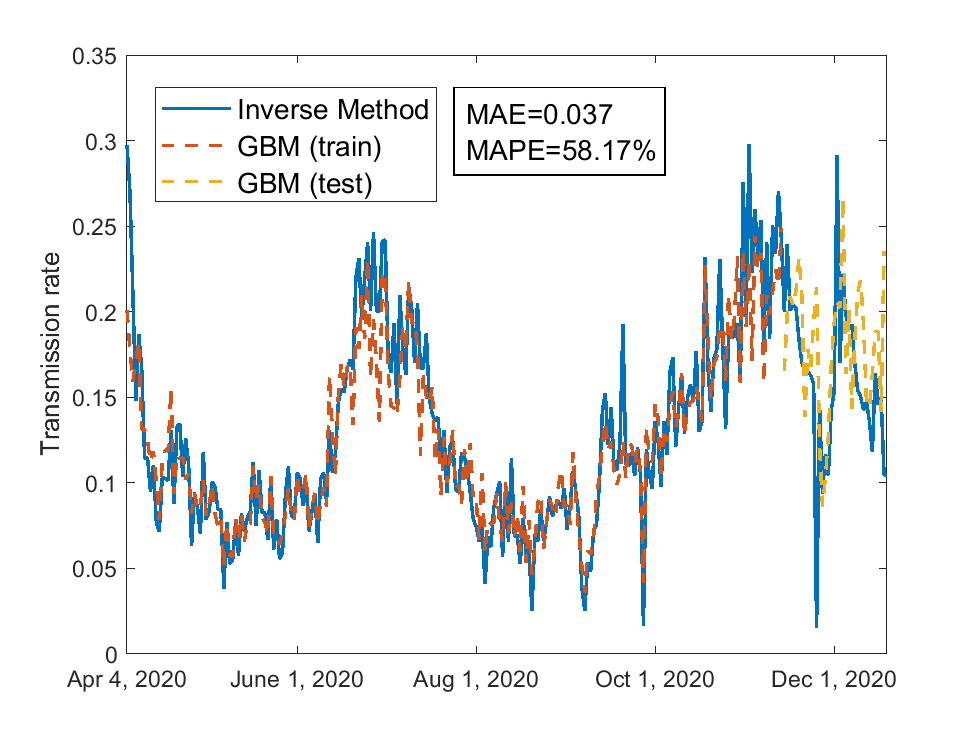}
\includegraphics[width=0.49\textwidth]{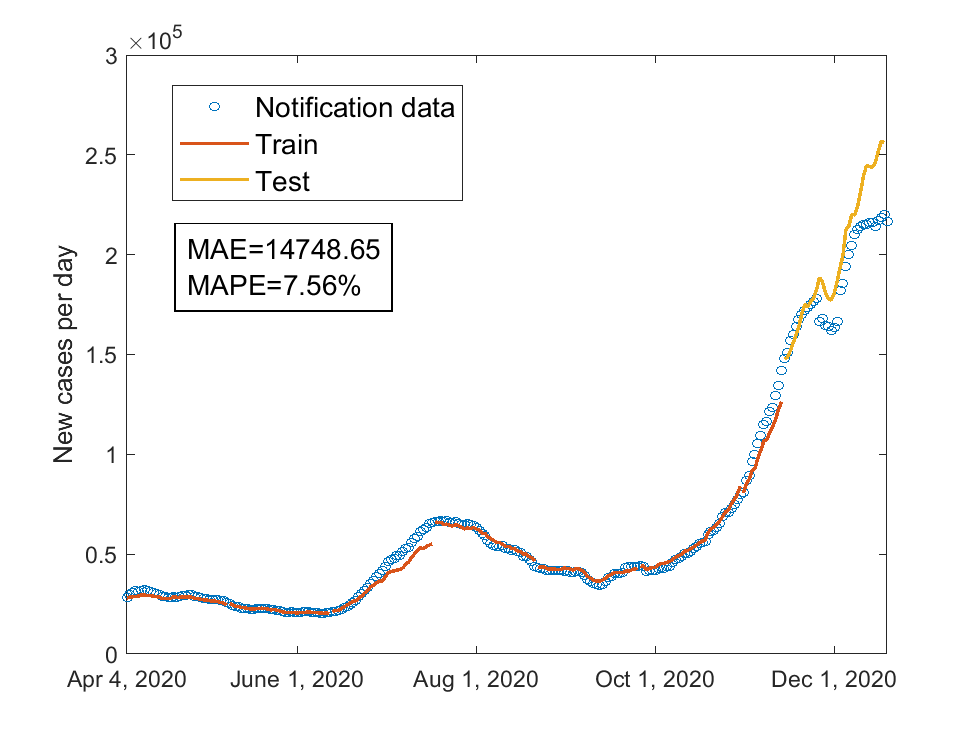}
\caption{\label{preFit1Mobility} Using mobility data M1$\sim$M6, train 224 days from Apr 4, 2020 to Nov 13, 2020; test 35 days from Nov 14, 2020 to Dec 18, 2020.}
\end{figure}

The prediction performance for the confirmed cases of
COVID-19 is summarized in Table \ref{MAEtable1}. 
We can see that the averaged MAE and MAPE of the GBM with both mobility and policy predictors are smaller than those of the GBM with mobility predictors only, which indicates that involving policy data can produce better prediction results. 
In particular, very small MAEs and MAPEs are obtained for the prediction results of daily confirmed cases when the GBM involving both mobility and policy variables is trained for 126 days, 133 days, and 217 days. The corresponding training and testing (prediction) results of the transmission rate together with the fitted curves of confirmed cases are presented in Figure \ref{fit1}, supplementary Figures \ref{fit2} and \ref{fit3}. 
The trained transmission rates (i.e., the orange curves in the left panels of these figures) generally fit well with the ones obtained from the inverse method (i.e., the blue curves in the left panels of these figures). However, the tested transmission rate (i.e., the yellow curves in the left panels of these figures) do not fit well with the peaks or troughs of the blue curves of the transmission rate. In the right panels of Figure \ref{fit1} and supplementary Figures \ref{fit2}, \ref{fit3}, the orange curves (i.e., trained part) fit almost perfectly with the real notification data of confirmed cases. The yellow curves of prediction in the right panels of Figure \ref{fit1} and supplementary Figure \ref{fit2} also fit quite well with the blue circles of real data, with the MAPE equal to $2.86\%$ and $4.66\%$, respectively. In the right panel of supplementary Figure \ref{fit3}, the yellow prediction curve does not show a good fitting with the local minimum point around November 30 although the MAPE is as small as $5.64\%$. This may be because it is near the Thanksgiving holiday during which people get together and may not have many testings as usual.
For the GBM which involves only mobility variables as predictors, smaller MAE and MAPE are obtained when the model is trained for 224 days as shown in Figure \ref{preFit1Mobility}. In this case, the predicted result is able to show the local minimum of daily confirmed cases around November 30 (see the yellow curve in the right panel of Figure \ref{preFit1Mobility}).

\begin{figure}[ht]
\centering
\includegraphics[width=0.6\textwidth]{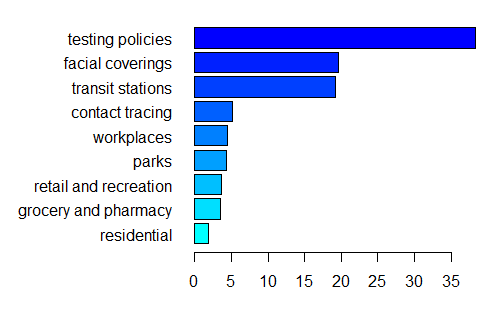}
\caption{\label{figPre126} Relative influence of mobility and H2, H3, H6 policy variables when trained for 126 days from Apr 4, 2020 to Aug 7, 2020.}
\end{figure}

\begin{figure}[ht]
\centering
\includegraphics[width=0.6\textwidth]{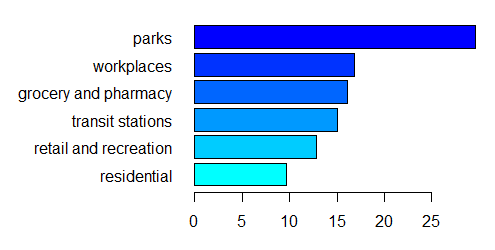}
\caption{\label{figPre224} Relative influence of mobility variables when trained for 224 days from Apr 4, 2020 to Nov 13, 2020.}
\end{figure}

The relative influence of the variables for the GBM involving both mobility and policy based on the different training durations of $126$ days, $133$ days, and $217$ days are shown in Table \ref{tablePre126}, supplementary Tables
\ref{tablePre133}, \ref{tablePre217},
and Figure \ref{figPre126},
supplementary Figures \ref{figPre133},
\ref{figPre217}, respectively. 
Among the three policies, the testing policy H2 always has the highest weight of relative influence which is as large as $38.37\%$ when trained for $126$ days. The second most important predictor is the facial covering policy H6 which weighs from about $18.36\%$ to $21.50\%$ corresponding to the above three training durations. The contact tracing policy H3 is the least important, with a weight ranging from about $5.09\%$ to $13.64\%$. 
As can be seen from Figure \ref{figPre126} and supplementary Figures \ref{figPre133}, \ref{figPre217}, the rankings of the relative influence of the mobility and policy variables have changed when trained for different lengths of days. When the policy variables are dropped, the ranking of the relative influence of mobility variables in Figure \ref{figPre224} is also different from those in Figure \ref{figPre126}, supplementary Figures \ref{figPre133} and \ref{figPre217}.

\section{Discussion}\label{Discussion}

We proposed a new framework for making predictions, that is, a hybrid model combining a mechanistic SEIAR model and gradient boosting models (GBM) with policy and mobility variables as predictors. We created an inverse method to estimate the time-varying transmission rate of COVID-19. This inverse method allows us to directly deal with time series data of daily confirmed cases without needing to get a smooth curve of the notification data at first or to substitute the integral form of any compartmental variables as the authors did in \cite{Kong2015,Pollicott2012}, which greatly simplifies the process of deriving the transmission rate. Using the transmission rate obtained by the inverse method can give an almost perfect fit with the notification data, which obviously outcompetes the traditionally used method of least squares. 
The tree-based method used by GBM increases the accuracy of prediction by turning ``weak learners'' into ``strong learners'' in a gradual, additive and sequential way \cite{Friedman2001}. Both MAE and MAPE are used for evaluating the prediction performance of the GBMs on the transmission rate as well as the fitting result of the number of confirmed cases by the SEIAR model. The selected GBM is capable of capturing the correlation between the transmission rate obtained from the inverse method and the policy as well as mobility variables so that accurate predictions of daily confirmed cases are made based on the SEIAR model and notification data. The bar plots of relative influence show that the most important policy is always restrictions on gatherings.

The method presented in this paper for connecting policy/mobility and transmission rate is data-driven and hypothesis-free. 
This is different from some other methods such as the least-squares method where one needs to make simplifying assumptions on the form of the transmission rate in the future, e.g., it is constant, piecewise constant, or a combination of sigmoid functions \cite{Lopez2021,Balcha2020,Sahoo2020,Tatrai2020,Ianni2020,Choi2020,Pluchino2021,Zhou2020}. 
The least-squares method makes future predictions based on either a pre-assumed form of the transmission rate function with respect to time or the ``current''(i.e., using the transmission rate on the last day of the training set as the transmission rate on each day of the prediction period), whereas machine-learning models make predictions based on the ``past'' (i.e., the trained experience), and typically without making restrictively simplifying assumptions.
In particular, our hybrid model that is based on only preventive policies and/or mobilities, is trained on ``past'' data to link the policies/mobilities to transmission rate, and uses ``future'' data on policies/mobilities to estimate the future values of the transmission rate. 
Given a set of ``future" policy/mobility data for the 35-day test window, we can get corresponding predicted values of the transmission rate during that window. As such, the model can be used to compare the dynamics under different future NPIs or mobility trends.
Non-pharmaceutical preventive policies are often \emph{a priori} known and available for making predictions.
In situations where the data is unavailable regularly or contains missing values, Bayesian networks can be used instead of GBM \cite{ramazi2021exploiting}.

Strikingly 90\% of the world's data have been generated in the past several years, thus machine learning has become more efficient in making predictions; however, mechanistic models can provide the causality missing from machine-learning approaches \cite{Baker2018}. 
Our hybrid model could provide more reliable predictions, especially when future policies have dramatic changes and enough amount of data are provided for training. 
Logically, our method has similar accuracy as machine learning approaches, but the disease spread compartment of our method includes a mechanistic model that captures established epidemiological causal relationships between the disease variables. 
In addition, there is no need to compare our method with the least-squares method because the inverse method has perfect data fitting for transmissibility without making any assumptions.

Since machine learning requires sufficient amount of data in order to obtain effective training, our hybrid model may not be competent in making predictions in the initial stage of an epidemic/pandemic caused by a novel pathogen. 
In addition, in our model we simply assume that human individuals in the US are homogeneously mixed and obtain policy data by averaging the policy data over different states together with Washington D.C. weighted by their populations. 
Indeed,
different states or regions usually have different epidemic progress and different preventive and control policies.
Even within a small region, 
different people may have different immunity abilities and hence different recovery or death rates, etc. 
To incorporate the role of heterogeneity in disease transmission, we can either apply our model and method to different smaller regions with region-specific parameters and then compare the prediction results or develop a patchy ODE model or PDE model with nonlocal dispersal.
We can also divide the population into more compartments according to their ages, health states or activity levels such as in different exposed periods, hospitalized, quarantined, on travel, working in medical frontlines, etc. and assume parameter values to be group-specific accordingly.

Our method can be applied to the study of other infectious diseases or future newly emerged pandemics in early stages. 
It can identify the most influential variables in predicting the disease spread and predict disease dynamics under different policies, which may guide policy makers to design mitigation measures. 
Our next step is to apply the inverse method plus machine learning approach to make predictions on daily new cases for the post-vaccination period and uncover the role of vaccination policies in future pandemic waves.

\section*{Acknowledgements}
This work was funded by Alberta Innovates and Pfizer via project number RES0052027. 
XW gratefully acknowledges support from Research and Creative Activity (RCA) Grant awarded by College of Arts and Sciences at the University of Tennessee at Chattanooga.
HW gratefully acknowledges support from Natural Sciences and Engineering Research Council of Canada (NSERC) Discovery Grant RGPIN-2020-03911 and NSERC Accelerator Grant RGPAS-2020-00090. KN gratefully acknowledges support from National Institute for Mathematical Sciences (NIMS) grant funded by the Korean Government (NIMS-B21910000). 
MAL is a Canada Research Chair in Mathematical Biology and gratefully acknowledges support from NSERC Discovery Grant.

\vspace{1.5cm}

\begin{table}[ht]
\centering
\begin{tabular}{l|l|l}
Parameter & Interpretation & Value \\\hline
$\beta(t)$ & Transmission rate & see Figure \ref{BetaPreVac} \\
$N$ & Total population of US & 331,449,281 \\
$\theta_E$ & Relative transmissibility of exposed individuals & 0.1\\ 
$\theta_A$ & Relative transmissibility of asymptomatic individuals  & 0.5 \\
$1/\delta$ & Incubation period &  14 days\\
$p$ & Proportion of asymptomatic infections & 0.7 \\
$\mu(t)$ & Death rate of symptomatic infected individuals & see Figure \ref{PreVacDeathRate} \\
$r_I$ & Recovery rate of symptomatic infected individuals & 1/14 day$^{-1}$ \\
$r_A$ & Recovery rate of asymptomatic infected individuals & 1/14 day$^{-1}$
\end{tabular}
\caption{\label{preParameters}Parameter interpretation and values.}
\end{table}

\begin{table}[ht]
\centering
\begin{tabular}{l|l|l}
Train length (days) & Train duration & Test duration \\\hline
105 & Apr 4, 2020 to Jul 17, 2020 & Jul 18, 2020 to Aug 21, 2020 \\
112 & Apr 4, 2020 to Jul 24, 2020 & Jul 25, 2020 to Aug 28, 2020 \\
119 & Apr 4, 2020 to Jul 31, 2020 &  Aug 1, 2020 to Sept 4, 2020 \\
126 & Apr 4, 2020 to Aug 7, 2020 &  Aug 8, 2020 to Sept 11, 2020 \\
133 &  Apr 4, 2020 to Aug 14, 2020 & Aug 15, 2020 to Sept 18, 2020 \\
140 &  Apr 4, 2020 to Aug 21, 2020 & Aug 22, 2020 to Sept 25, 2020 \\
147 &  Apr 4, 2020 to Aug 28, 2020 & Aug 29, 2020 to Oct 2, 2020 \\
154 &  Apr 4, 2020 to Sept 4, 2020 & Sept 5, 2020 to Oct 9, 2020 \\
161 &  Apr 4, 2020 to Sept 11, 2020 & Sept 12, 2020 to Oct 16, 2020 \\
168 &  Apr 4, 2020 to Sept 18, 2020 & Sept 19, 2020 to Oct 23, 2020 \\
175 &  Apr 4, 2020 to Sept 25, 2020 & Sept 26, 2020 to Oct 30, 2020 \\
182 &  Apr 4, 2020 to Oct 2, 2020 & Oct 3, 2020 to Nov 6, 2020 \\
189 &  Apr 4, 2020 to Oct 9, 2020
 & Oct 10, 2020 to Nov 13, 2020 \\
196 & Apr 4, 2020 to Oct 16, 2020
& Oct 17, 2020 to Nov 20, 2020 \\
203 & Apr 4, 2020 to Oct 23, 2020
& Oct 24, 2020 to Nov 27, 2020 \\
210 & Apr 4, 2020 to Oct 30, 2020
& Oct 31, 2020 to Dec 4, 2020 \\
217 & Apr 4, 2020 to Nov 6, 2020
& Nov 7, 2020 to Dec 11, 2020 \\
224 & Apr 4, 2020 to Nov 13, 2020
& Nov 14, 2020 to Dec 18, 2020 
\end{tabular}
\caption{\label{preVac}Training and testing durations.}
\end{table}

\begin{table}[ht]
\centering
\begin{tabular}{l|l|l|l}
Train length (days) & Train duration & MAE & MAPE ($\%$) \\\hline
105 & Apr 4, 2020 to Jul 17, 2020 &36616.64 & 70.07\\
112 & Apr 4, 2020 to Jul 24, 2020 &27423.54  & 58.03\\
119 & Apr 4, 2020 to Jul 31, 2020 &17786.35   & 39.88\\
126 & Apr 4, 2020 to Aug 7, 2020 & 2059.83  & 4.92\\
133 &  Apr 4, 2020 to Aug 14, 2020 & 6063.13 & 15.61\\
140 &  Apr 4, 2020 to Aug 21, 2020 & 5714.62 & 13.85\\
147 &  Apr 4, 2020 to Aug 28, 2020 & 2451.90 & 6.47\\
154 &  Apr 4, 2020 to Sept 4, 2020 & 2540.38 & 6.45\\
161 &  Apr 4, 2020 to Sept 11, 2020 & 11753.02 & 26.39\\
168 &  Apr 4, 2020 to Sept 18, 2020 & 9150.12 & 17.76\\
175 &  Apr 4, 2020 to Sept 25, 2020 & 4268.39 & 8.53\\
182 &  Apr 4, 2020 to Oct 2, 2020 & 19195.41 & 24.87\\
189 &  Apr 4, 2020 to Oct 9, 2020
 & 25871.46 & 25.30\\
196 & Apr 4, 2020 to Oct 16, 2020
& 41907.22 &33.51\\
203 & Apr 4, 2020 to Oct 23, 2020
& 35601.46 & 23.23\\
210 & Apr 4, 2020 to Oct 30, 2020
& 59973.39 & 37.70\\
217 & Apr 4, 2020 to Nov 6, 2020
& 56162.49 & 30.63\\
224 & Apr 4, 2020 to Nov 13, 2020
& 70696.84 & 36.14
\end{tabular}
\caption{\label{allMAEMAPE}MAE and MAPE of predictions of notification data based on model \eqref{pre_model} and the GBM in Section \ref{MachineLearningPolicySection} corresponding to different training durations.}
\end{table}

\begin{figure}[ht]
\centering
\includegraphics[width=0.49\textwidth]{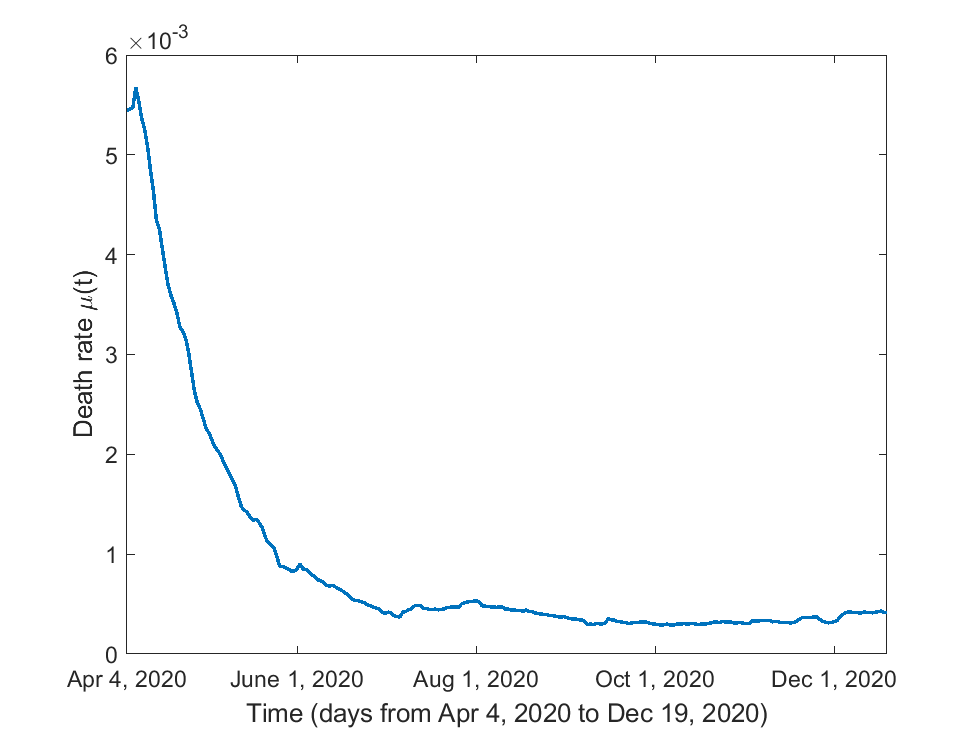}
\caption{\label{PreVacDeathRate} Disease induced death rate from Apr 4, 2020 to Dec 19, 2020.}
\end{figure}

\begin{figure}[ht]
\centering
\includegraphics[width=0.49\textwidth]{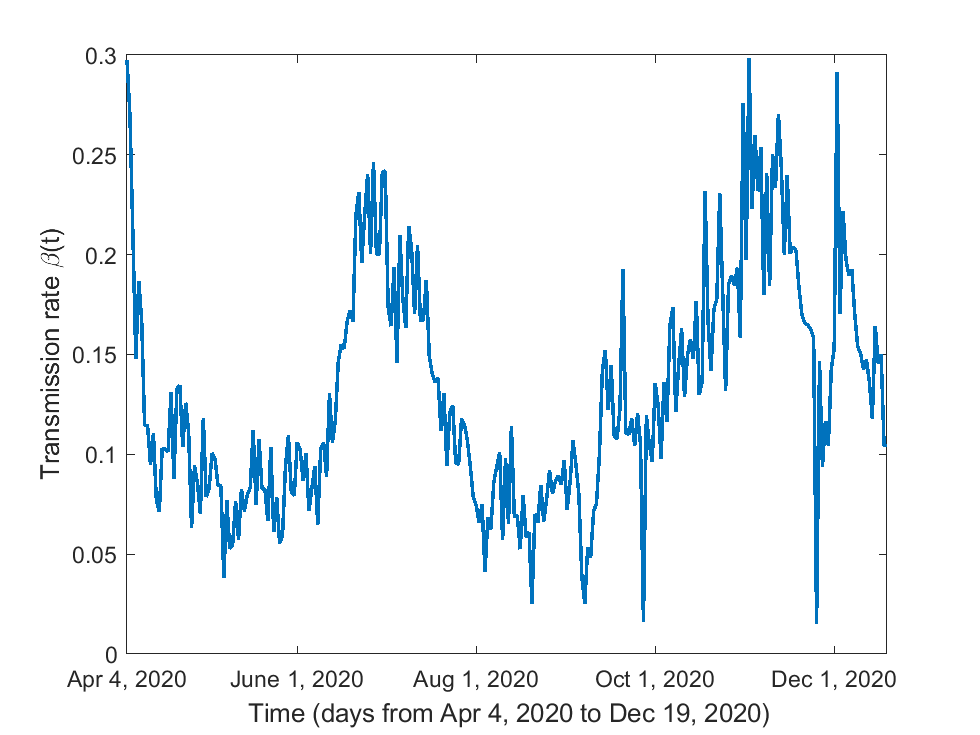}
\includegraphics[width=0.49\textwidth]{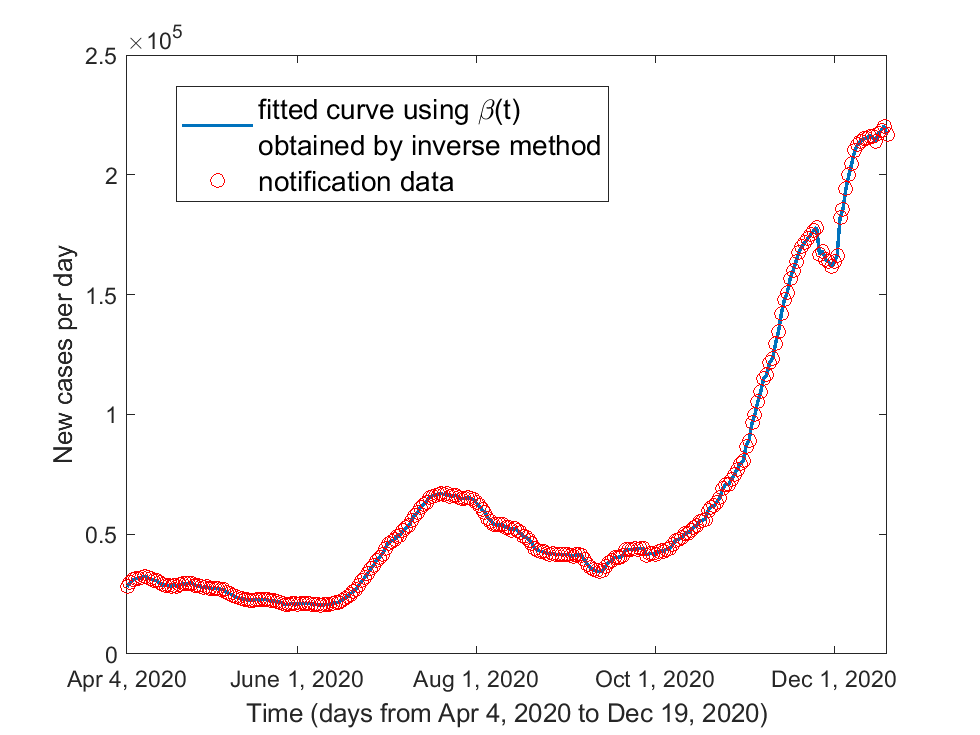}
\caption{\label{BetaPreVac} Transmission rate obtained by the inverse method and the fitting with notification data from Apr 4, 2020 to Dec 19, 2020.}
\end{figure}

\begin{table}[ht]
\centering
\begin{tabular}{l|l|l}
Data used in GBM & Averaged MAE & Averaged MAPE \\\hline
Policy data C1$\sim$C8, H1, H2, H3, H6, H8
& $24179.79$
& $26.63\%$
\end{tabular}
\caption{\label{averMAEMAPE}Averaged MAE and MAPE for the prediction of daily confirmed cases by using model \eqref{pre_model} and the GBM in Section \ref{MachineLearningPolicySection}.}
\end{table}

\begin{figure}[ht]
\centering
\includegraphics[width=0.49\textwidth]{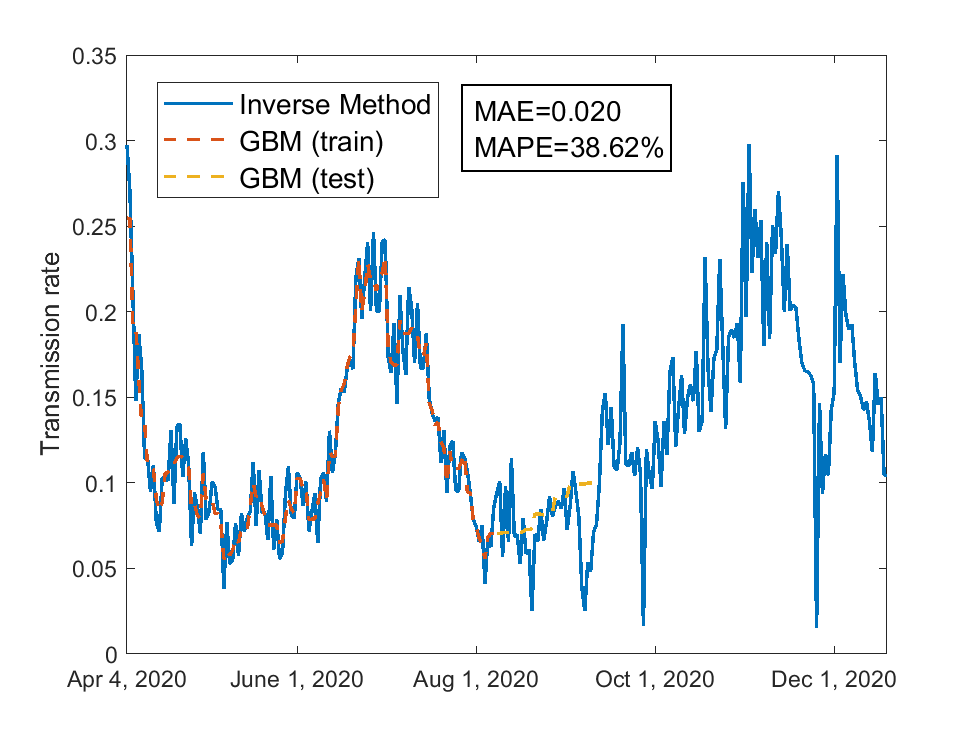}
\includegraphics[width=0.49\textwidth]{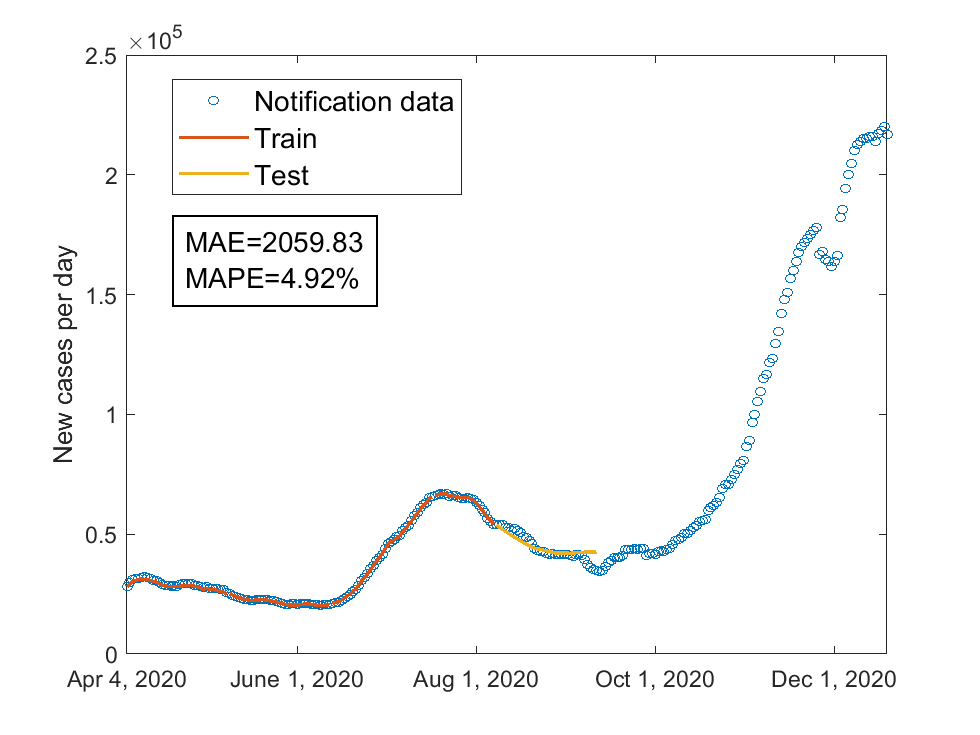}
\caption{\label{fit01} Using policy data C1$\sim$C8, H1, H2, H3, H6 and H8, train 126 days from Apr 4, 2020 to Aug 7, 2020; test 35 days from Aug 8, 2020 to Sept 11, 2020.}
\end{figure}

\begin{table}[ht]
\centering
\begin{tabular}{l|l}
Variable & Relative influence ($\%$)\\\hline
restrictions on gatherings &       31.0489157\\
testing policies &                 16.7287051\\
protection of elderly people&      15.2217062\\
school closing &                    7.9329945\\
facial coverings  &                 6.0243431\\
workplace closing &                 5.0720110\\
restrictions on internal movement & 4.5464197\\
close public transport  &           4.0595600\\
cancel public events  &             3.4738417\\
stay at home requirements &         3.2881507\\
contact tracing  &                  2.4412499\\
public information campaigns &      0.1621026\\
international travel controls &     0.0000000
\end{tabular}
\caption{\label{tablePre126P} Relative influence of policy variables when trained for 126 days from Apr 4, 2020 to Aug 7, 2020.}
\end{table}

\begin{table}[ht]
\centering
\begin{tabular}{l|l|l}
Data used in GBM & Averaged MAE & Averaged MAPE \\\hline
Mobility data
& $26188.58$
& $36.22\%$\\
Mobility data $+$ policy data H2, H3, H6
& $20408.11$
& $25.67\%$
\end{tabular}
\caption{\label{MAEtable1}Averaged MAE and MAPE for the prediction of daily confirmed cases by using model \eqref{pre_model}+the GBM with mobility only and model \eqref{pre_model}+the GBM with mobility and policy as predictors.}
\end{table}

\begin{table}[ht]
\centering
\begin{tabular}{l|l}
Variable & Relative influence ($\%$)\\\hline
testing policies &38.370821\\
facial coverings &19.574882\\
transit stations &19.215391\\
contact tracing  &5.093295\\
workplaces  &4.396082\\
parks  &4.316401\\
retail and recreation & 3.614830\\
grocery and pharmacy & 3.556656\\
residential & 1.861642
\end{tabular}
\caption{\label{tablePre126} Relative influence of mobility and H2, H3, H6 policy variables when trained for 126 days from Apr 4, 2020 to Aug 7, 2020.}
\end{table}

%\begin{figure}[ht]
%\centering
%\includegraphics[width=0.32\textwidth]{preBetaH2.png}
%\includegraphics[width=0.32\textwidth]{preBetaH3.png}
%\includegraphics[width=0.32\textwidth]{preBetaH6.png}
%\includegraphics[width=0.32\textwidth]{preBetaM1.png}
%\includegraphics[width=0.32\textwidth]{preBetaM2.png}
%\includegraphics[width=0.32\textwidth]{preBetaM3.png}
%\includegraphics[width=0.32\textwidth]{preBetaM4.png}
%\includegraphics[width=0.32\textwidth]{preBetaM5.png}
%\includegraphics[width=0.32\textwidth]{preBetaM6.png}
%\caption{\label{prePDP} Partial dependence plots based on the training from Apr 4, 2020 to Nov 6, 2020.}
%\end{figure}

%\begin{figure}[ht]
%\centering
%\includegraphics[width=0.32\textwidth]{preH2H3.png}
%\includegraphics[width=0.32\textwidth]{preH2H6.png}
%\includegraphics[width=0.32\textwidth]{preH3H6.png}
%\includegraphics[width=0.32\textwidth]{CpreH2H3.png}
%\includegraphics[width=0.32\textwidth]{CpreH2H6.png}
%\includegraphics[width=0.32\textwidth]{CpreH3H6.png}
%\caption{\label{CprePDP} 2D partial dependence plots based on the training from Apr 4, 2020 to Nov 6, 2020.}
%\end{figure}

\begin{table}[ht]
\centering
\begin{tabular}{l|l}
Variable & Relative influence ($\%$)\\\hline
parks  &                             29.631102\\
workplaces  &                   16.835859\\
grocery and pharmacy  & 16.086562\\
transit stations  &              14.995899\\
retail and recreation  &      12.799867\\
residential    &                   9.650711\\
\end{tabular}
\caption{\label{tablePre224} Relative influence of mobility variables when trained for 224 days from Apr 4, 2020 to Nov 13, 2020.}
\end{table}

\clearpage
\bibliographystyle{alpha}
\bibliography{sample}

\bigskip

\appendix

\section*{Appendix. Supplementary figures and tables}
\begin{figure}[ht]
\centering
\includegraphics[width=0.49\textwidth]{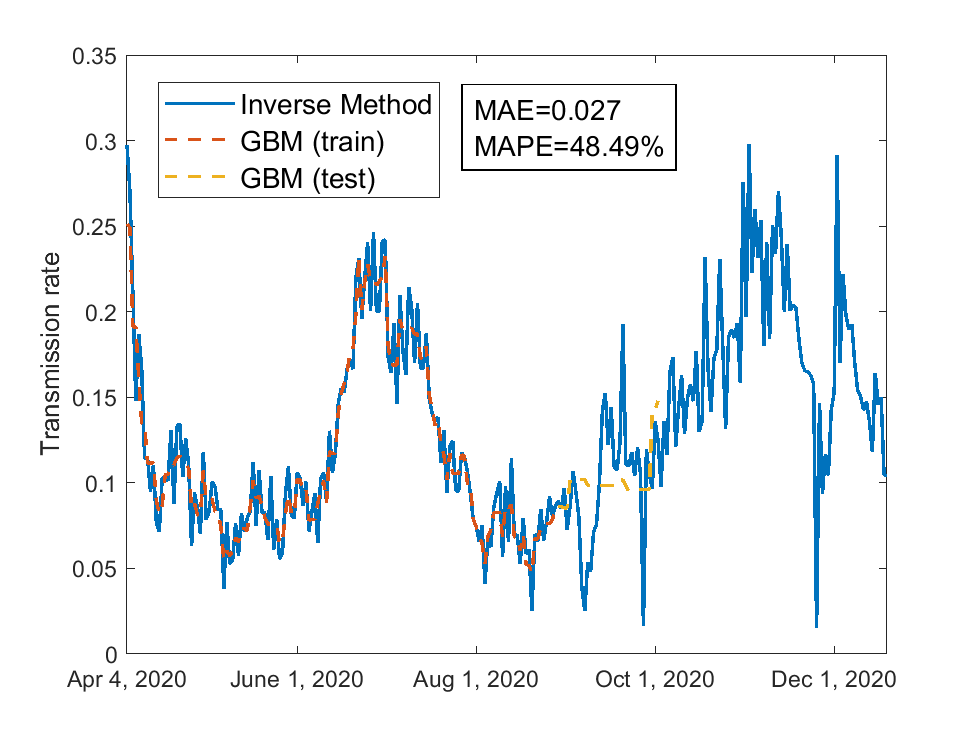}
\includegraphics[width=0.49\textwidth]{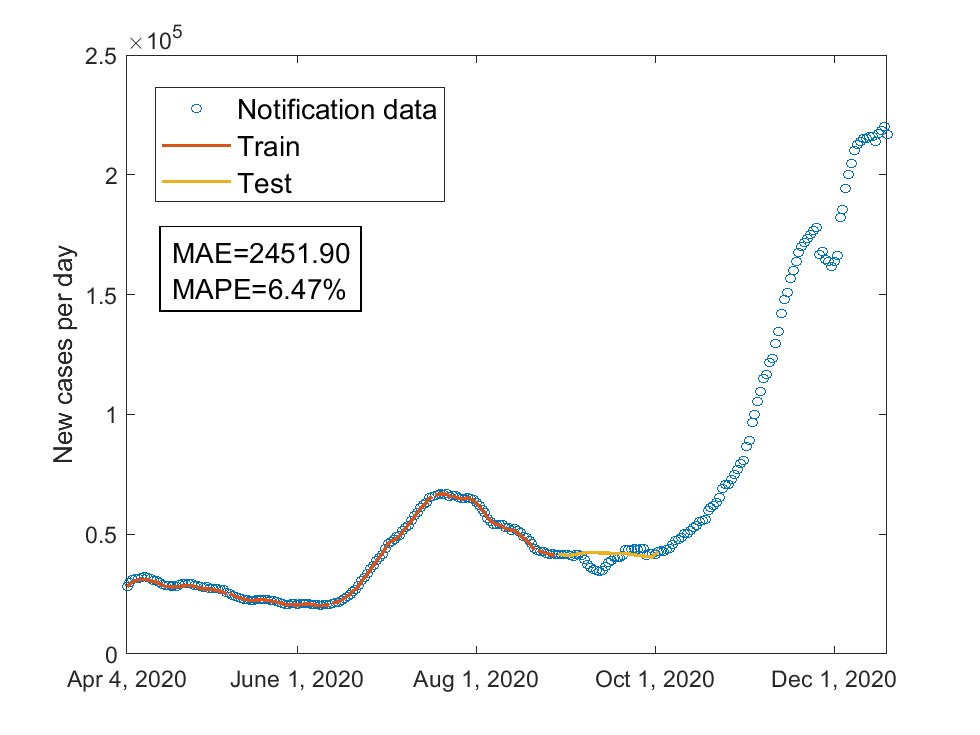}
\caption{\label{fit02} Using policy data C1$\sim$C8, H1, H2, H3, H6 and H8, train 147 days from Apr 4, 2020 to Aug 28, 2020; test 35 days from Aug 29, 2020 to Oct 2, 2020.}
\end{figure}

\begin{figure}[ht]
\centering
\includegraphics[width=0.6\textwidth]{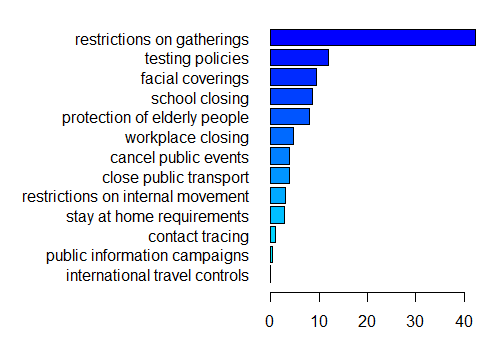}
\caption{\label{figPre147P} Relative influence of policy variables when trained for 147 days from Apr 4, 2020 to Aug 28, 2020.}
\end{figure}

\begin{figure}[ht]
\centering
\includegraphics[width=0.49\textwidth]{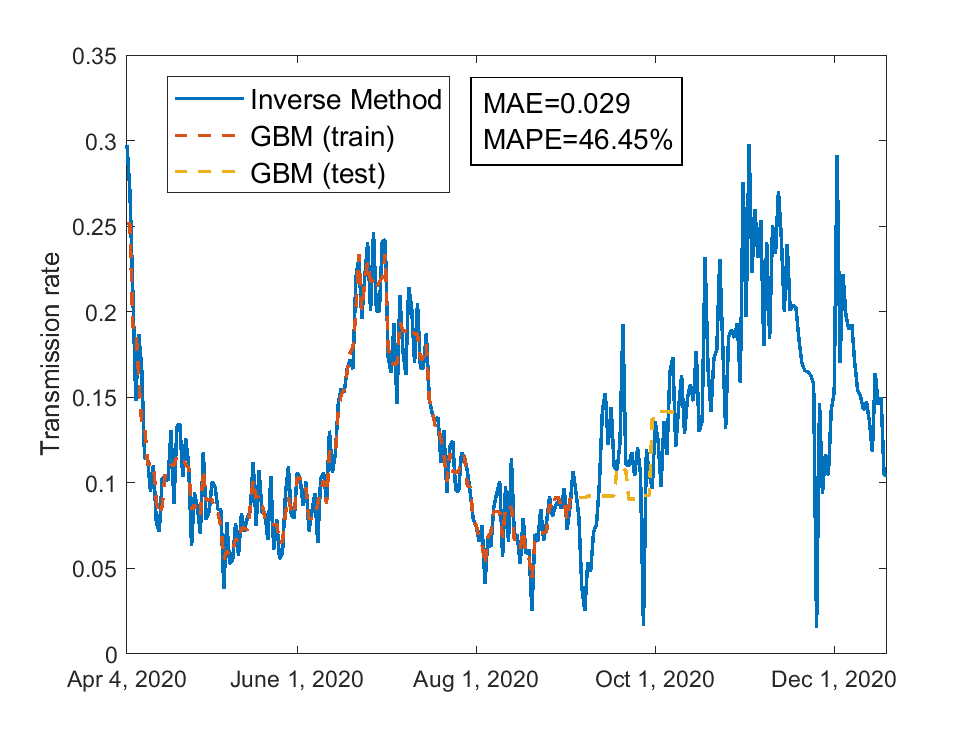}
\includegraphics[width=0.49\textwidth]{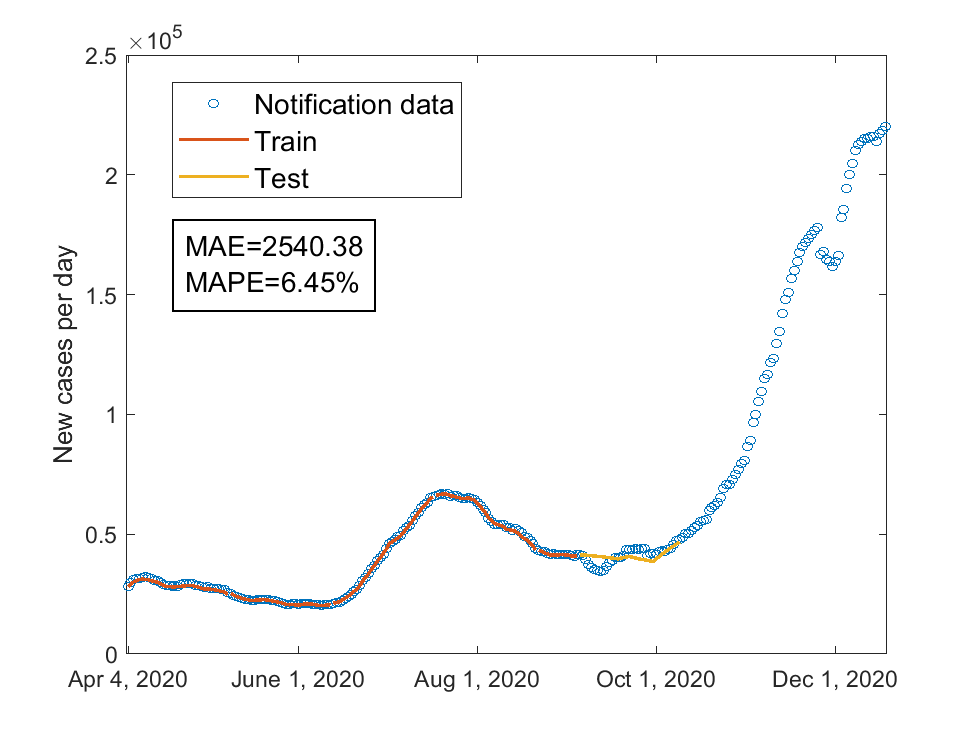}
\caption{\label{fit03} Using policy data C1$\sim$C8, H1, H2, H3, H6 and H8, train 154 days from Apr 4, 2020 to Sept 4, 2020; test 35 days from Sept 5, 2020 to Oct 9, 2020.}
\end{figure}

\begin{figure}[ht]
\centering
\includegraphics[width=0.6\textwidth]{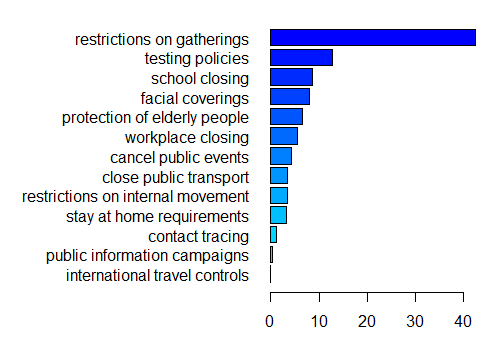}
\caption{\label{figPre154P} Relative influence of policy variables when trained for 154 days from Apr 4, 2020 to Sept 4, 2020.}
\end{figure}

\begin{figure}[ht]
\centering
\includegraphics[width=0.49\textwidth]{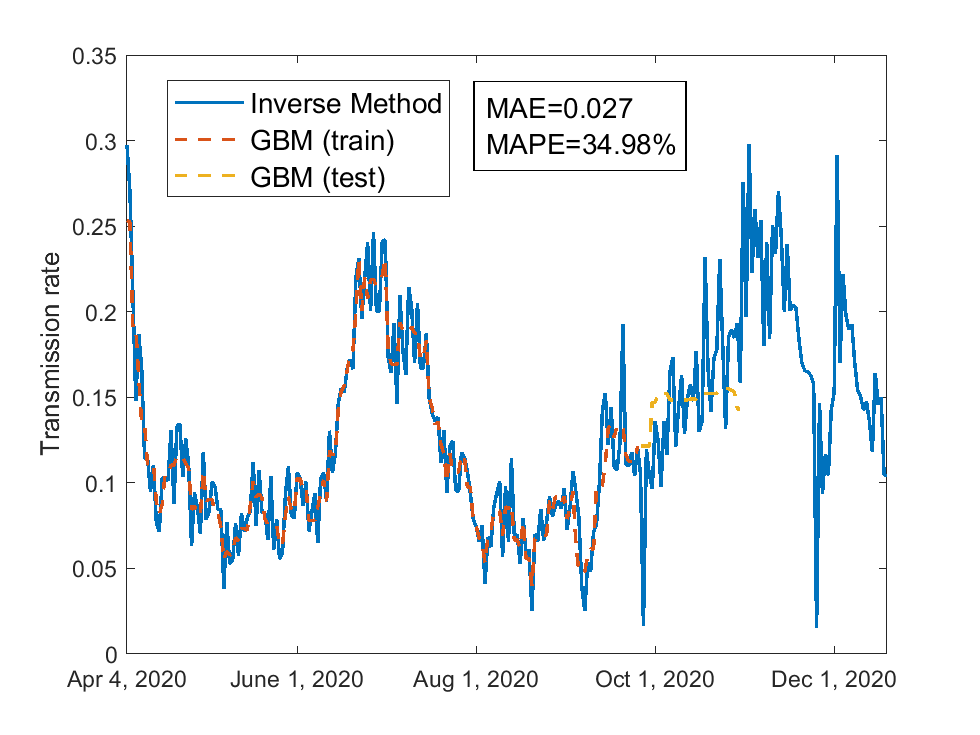}
\includegraphics[width=0.49\textwidth]{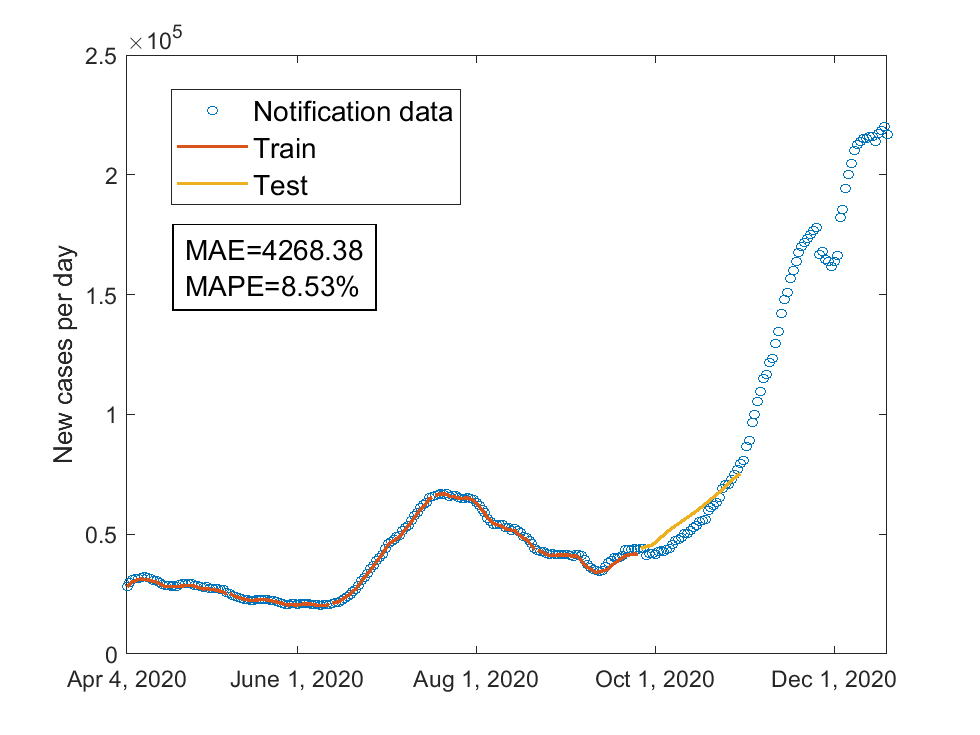}
\caption{\label{fit04} Using policy data C1$\sim$C8, H1, H2, H3, H6 and H8, train 175 days from Apr 4, 2020 to Sept 25, 2020; test 35 days from Sept 26, 2020 to Oct 30, 2020.}
\end{figure}

\begin{figure}[ht]
\centering
\includegraphics[width=0.6\textwidth]{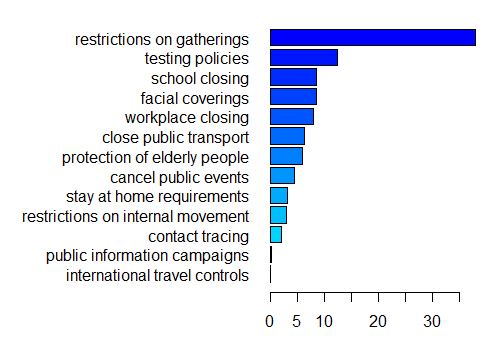}
\caption{\label{figPre175P} Relative influence of policy variables when trained for 175 days from Apr 4, 2020 to Sept 25, 2020.}
\end{figure}

\begin{figure}[ht]
\centering
\includegraphics[width=0.49\textwidth]{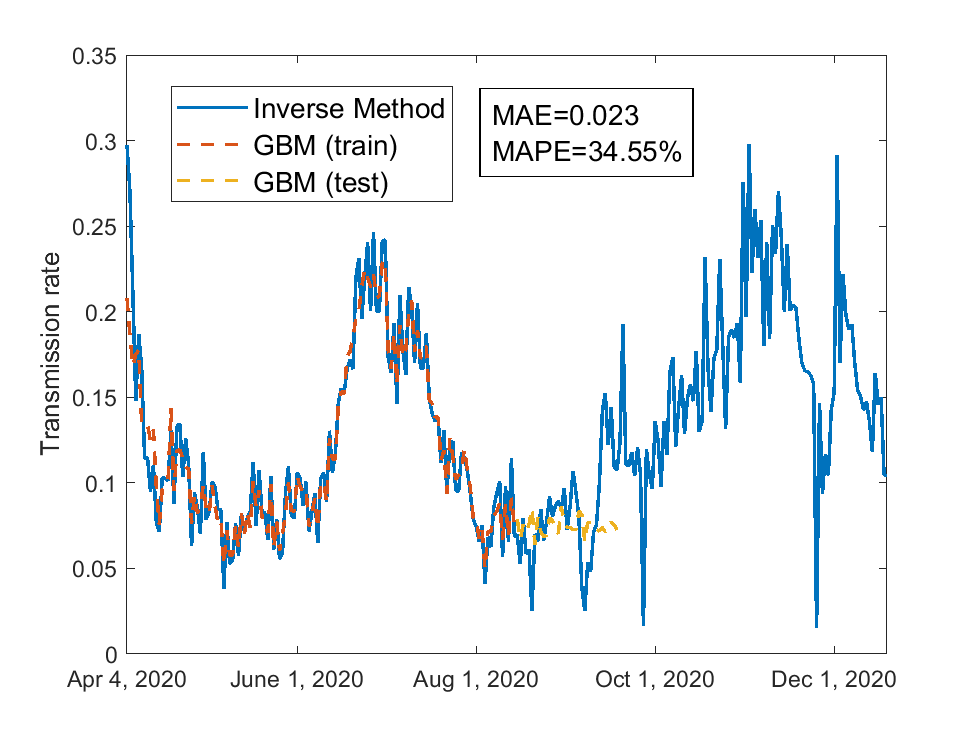}
\includegraphics[width=0.49\textwidth]{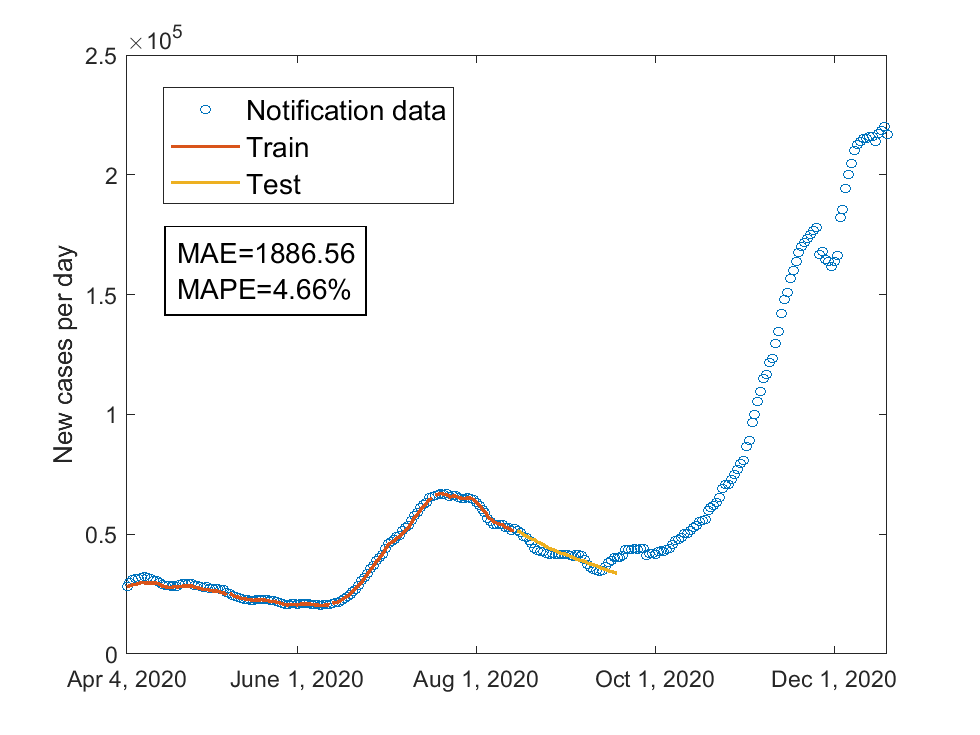}
\caption{\label{fit2} Using mobility data and H2, H3, H6 policy data, train 133 days from Apr 4, 2020 to Aug 14, 2020; test 35 days from Aug 15, 2020 to Sept 18, 2020.}
\end{figure}

\begin{figure}[ht]
\centering
\includegraphics[width=0.6\textwidth]{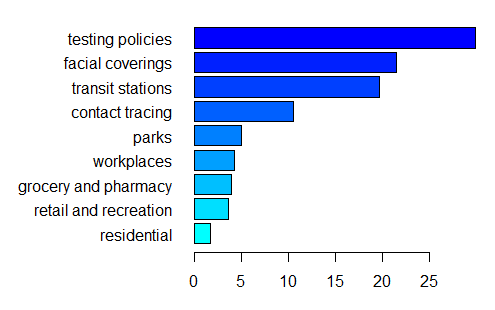}
\caption{\label{figPre133} Relative influence of mobility and H2, H3, H6 policy variables when trained for 133 days from Apr 4, 2020 to Aug 14, 2020.}
\end{figure}

\begin{figure}[ht]
\centering
\includegraphics[width=0.49\textwidth]{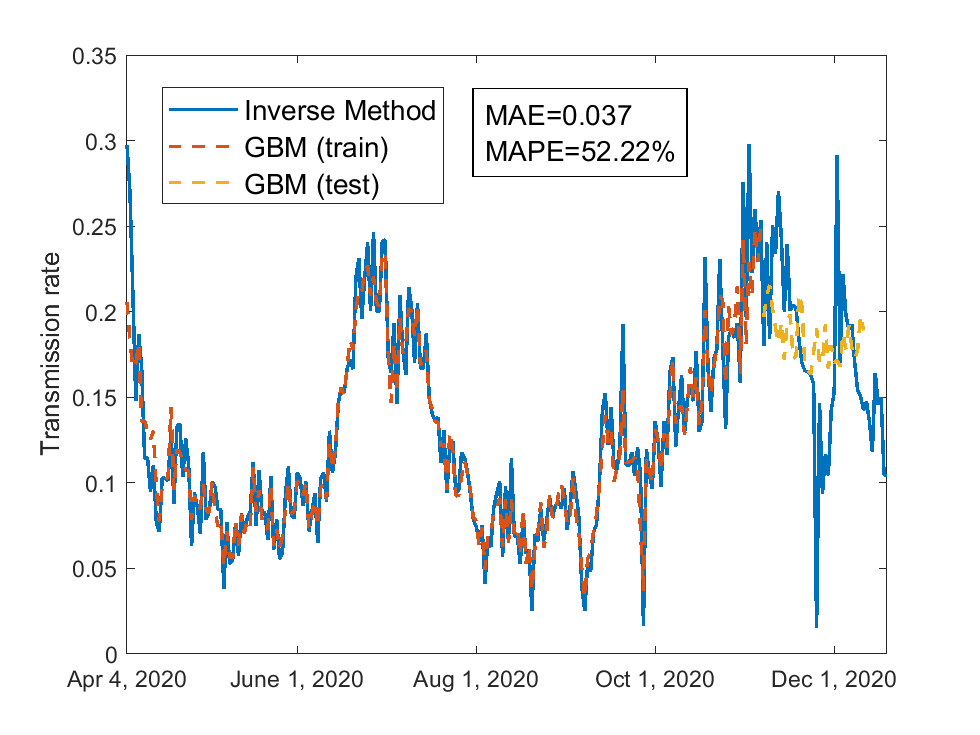}
\includegraphics[width=0.49\textwidth]{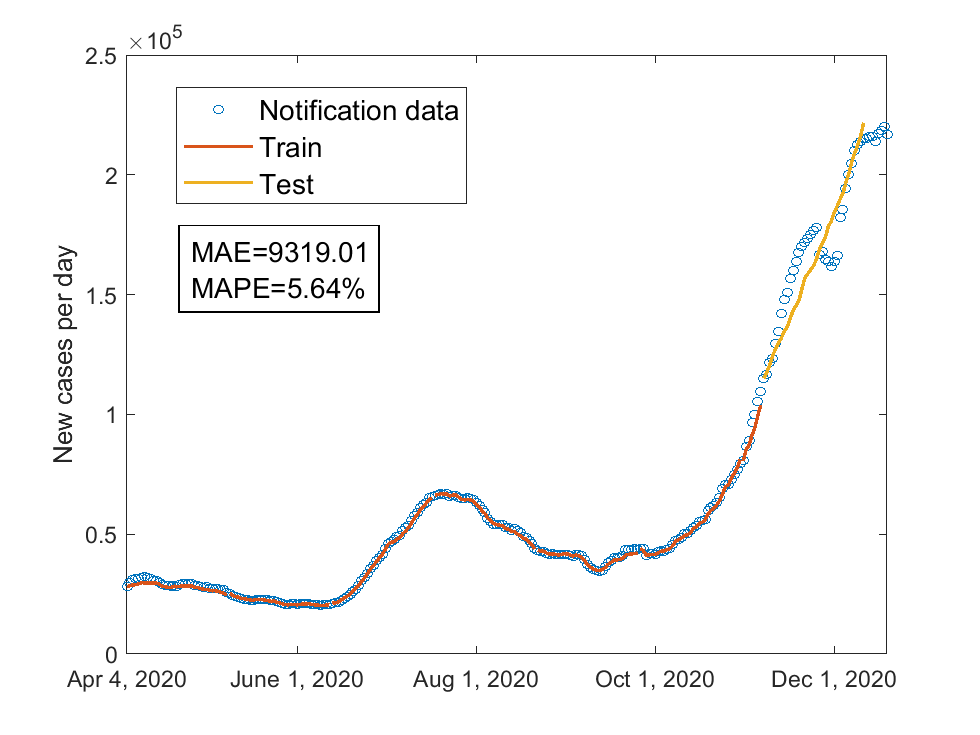}
\caption{\label{fit3} Using mobility data and H2, H3, H6 policy data, train 217 days from Apr 4, 2020 to Nov 6, 2020; test 35 days from Nov 7, 2020 to Dec 11, 2020.}
\end{figure}

\begin{figure}[ht]
\centering
\includegraphics[width=0.6\textwidth]{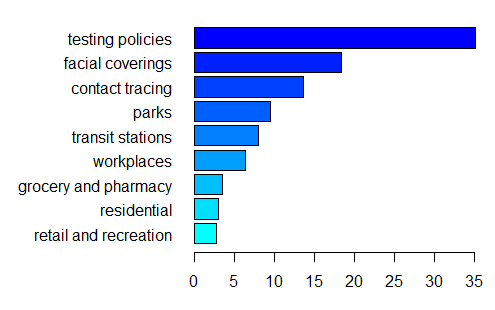}
\caption{\label{figPre217} Relative influence of mobility and H2, H3, H6 policy variables when trained for 217 days from Apr 4, 2020 to Nov 6, 2020.}
\end{figure}

In this Appendix, we present supplementary figures and tables. 
The selected training and testing results about the transmission rates and the fittings with notification data of daily confirmed cases are displayed in Figures 
\ref{fit02}, \ref{fit03}, \ref{fit04}
for the model with policy as the only predictors, in Figures 
\ref{fit2}, \ref{fit3} for the model with both policy and mobility as the predictors.
After each of these figures we present a table and a figure of the relative influence of the involved predictor variables in training the model.
Tables \ref{tablePre147P}, \ref{tablePre154P}, \ref{tablePre175P} and Figures \ref{figPre147P}, \ref{figPre154P}, \ref{figPre175P} show the relative influence of the policy vairables when the model is trained for 147 days, 154 days, 175 days, respectively.
Tables \ref{tablePre133}, \ref{tablePre217} and  Figures \ref{figPre133}, \ref{figPre217}
give the relative influence of the mobility and part of policy variables when the model is trained for 133 days and 217 days, respectively.

\begin{table}[ht]
\centering
\begin{tabular}{l|l}
Variable & Relative influence ($\%$)\\\hline
restrictions on gatherings &       42.3668084\\
testing policies &                 11.9967018\\
facial coverings &                  9.4461173\\
school closing  &                   8.7000103\\
protection of elderly people &      7.8984697\\
workplace closing &                 4.6655947\\
cancel public events  &             3.9032770\\
close public transport &            3.7699113\\
restrictions on internal movement & 3.1179248\\
stay at home requirements &         2.8130527\\
contact tracing  &                  0.8900337\\
public information campaigns&      0.4320982\\
international travel controls&      0.0000000
\end{tabular}
\caption{\label{tablePre147P} Relative influence of policy variables when trained for 147 days from Apr 4, 2020 to Aug 28, 2020.}
\end{table}

\begin{table}[ht]
\centering
\begin{tabular}{l|l}
Variable & Relative influence ($\%$)\\\hline
restrictions on gatherings &       42.4561382\\
testing policies &                 12.8734097\\
school closing &                    8.5351249\\
facial coverings &                  8.0608618\\
protection of elderly people &      6.6531206\\
workplace closing &                 5.6067289\\
cancel public events &              4.3324652\\
close public transport &            3.4850619\\
restrictions on internal movement & 3.4346518\\
stay at home requirements &         3.1492062\\
contact tracing &                   1.0979493\\
public information campaigns &      0.3152815\\
international travel controls&      0.0000000
\end{tabular}
\caption{\label{tablePre154P} Relative influence of policy variables when trained for 154 days from Apr 4, 2020 to Sept 4, 2020.}
\end{table}

\begin{table}[ht]
\centering
\begin{tabular}{l|l}
Variable & Relative influence ($\%$)\\\hline
restrictions on gatherings &       38.0716839\\
testing policies &                 12.3025438\\
school closing &                    8.4113758\\
facial coverings &                  8.4021886\\
workplace closing &                 7.9880485\\
close public transport &            6.1772695\\
protection of elderly people &      5.9374740\\
cancel public events &              4.4452982\\
stay at home requirements &         3.1632396\\
restrictions on internal movement & 2.9548620\\
contact tracing &                   1.9726367\\
public information campaigns &      0.1733794\\
international travel controls&      0.0000000
\end{tabular}
\caption{\label{tablePre175P} Relative influence of policy variables when trained for 175 days from Apr 4, 2020 to Sept 25, 2020.}
\end{table}

\begin{table}[ht]
\centering
\begin{tabular}{l|l}
Variable & Relative influence ($\%$)\\\hline
testing policies &29.935771\\
facial coverings& 21.503403\\
transit stations &19.706089\\
contact tracing &10.468970\\
parks & 4.930258\\
workplaces & 4.243045\\
grocery and pharmacy & 3.940721\\
retail and recreation & 3.599258\\
residential & 1.672485
\end{tabular}
\caption{\label{tablePre133} Relative influence of mobility and H2, H3, H6 policy variables when trained for 133 days from Apr 4, 2020 to Aug 14, 2020.}
\end{table}

\begin{table}[ht]
\centering
\begin{tabular}{l|l}
Variable & Relative influence ($\%$) \\\hline
testing policies &35.148781\\
facial coverings &18.359430\\
contact tracing &13.641922\\
parks & 9.487232\\
transit stations & 7.922605\\
workplaces&  6.320635\\
grocery and pharmacy& 3.452438\\
residential & 3.007597\\
retail and recreation & 2.659359
\end{tabular}
\caption{\label{tablePre217} Relative influence of mobility and H2, H3, H6 policy variables when trained for 217 days from Apr 4, 2020 to Nov 6, 2020.}
\end{table}

\end{document}